\documentclass[12pt]{iopart}
\usepackage{graphicx,amsfonts,amsgen,amsbsy,amssymb,bbold,bm,cite}

\makeatletter
\def\@fnsymbol#1{^{\thefootnote}\relax}
\makeatother

\begin{document}

\article{Perspective}{Ferromagnetic Ferroelectricity due to Orbital Ordering}

\author{I V Solovyev}

\address{Research Center for Materials Nanoarchitectonics (MANA), National Institute for Materials Science (NIMS), 1-1 Namiki, Tsukuba, Ibaraki 305-0044, Japan}
\ead{SOLOVYEV.Igor@nims.go.jp}
\vspace{10pt}

\begin{abstract}
Realization of ferromagnetic ferroelectricity, combining two ferroic orders in a single phase, is the longstanding problem of great practical importance. One of the difficulties is that ferromagnetism alone cannot break inversion symmetry $\mathcal{I}$. Therefore, such a phase cannon be obtained by purely magnetic means. Here, we show how it can be designed by making orbital degrees of freedom active. The idea can be traced back to a basic principle of interatomic exchange, which states that an alternation of occupied orbitals along a bond (i.e., antiferro orbital order) favors ferromagnetic coupling. Moreover, the antiferro orbital order breaks $\mathcal{I}$, so that the bond becomes not simply ferromagnetic but also ferroelectric. Then, we formulate main principles governing the realization of such a state in solids, namely: (i) The magnetic atoms should not be located in inversion centers, as in the honeycomb lattice; (ii) The orbitals should be flexible enough to adjust they shape and minimize the energy of exchange interactions; (iii) This flexibility can be achieved by intraatomic interactions, which are responsible for Hund's second rule and compete with the crystal field splitting; (iv) For octahedrally coordinated transition-metal compounds, the most promising candidates appear to be iodides with a $d^{2}$ configuration and relatively weak $d$–$p$ hybridization. Such a situation is realized in the van der Walls compound VI$_3$, which we expect to be ferromagnetic ferroelectric. 
\end{abstract}

%
\vspace{2pc}
\par \submitto{\JPCM}
%
%
%

\section{\label{sec:Intro} Introduction}
\par Multiferroicity means a coexistence of two or more ferroic orders within one phase. Thus, literally, it should be a combination of \emph{ferro}magnetism, \emph{ferro}electricity, \emph{ferro}elasticity, or any other property with the prefix \emph{ferro}. Such a combination is of great practical importance: Because these ferroic properties are often intertwined, multiferroics offer an excellent platform for cross-control of polarization by the magnetic field and magnetization by the electric field~\cite{TokuraScience}. The prefix \emph{ferro} appears to be very important also in this context: the larger the magnetization (polarization), the weaker the magnetic (electric) field needed to achieve this cross-control. 

\par However, today the term multiferroicity is also understood in a more general sense, when ferroelectricity coexists with \emph{any} type of magnetic order and not necessarily the ferromagnetic (FM) one~\cite{CheongMostovoy,Khomskii2009,TokuraSekiNagaosa}, while ferromagnetic ferroelectrics are very rare~\cite{Hill}.

\par Ferromagnetic ferroelectricity is expected in materials with distinct sublattices, where one sublattice hosts ferromagnetism and another ferroelectricity~\cite{SeshadriHill,FennieRabe}. Typically, these ferroic properties have different microscopic origin and only weakly depend on each other. A notable exception is SrMnO$_3$ under epitaxial strain, which was predicted to be simultaneously ferromagnetic and ferroelectric (FE), and both of these properties stem from the Mn sublattice~\cite{LeeRabe,EdstromEderer}. Another possibility is the synthesis of artificial heterostructures, combining layers of FM and FE materials~\cite{heterostructures}.  

\par A new route for realizing ferromagnetic ferroelectricity has been proposed in~\cite{PRB2024}. The basic idea is traced back to Goodenough-Kanamori-Anderson (GKA) rules for interatomic exchange interactions~\cite{Anderson1950,Goodenough1955,Goodenough1958,Kanamori1959}, which state basically that population of alike orbitals at two sites of the bond (the ferro orbital order) favors antiferromagnetic (AFM) coupling, while population of unlike orbitals (the antiferro orbital order) will make this coupling ferromagnetic. What was overlooked in the canonical GKA picture is that, besides FM coupling, the antiferro orbital order breaks inversion symmetry $\mathcal{I}$ in the bond so that it becomes simultaneously ferromagnetic and ferroelectric. The idea was further adapted for periodic solids, arguing that in certain materials the spontaneous antiferro orbital order can yield the FM-FE state. The van der Walls FM semiconductor VI$_3$ is one of potential candidates for realizing such a state~\cite{PRB2024,ChDay}. 

\par In this article, we further elaborate basic conditions and perspectives of realizing the orbitally induced FM-FE state in real compounds. After reminding in \Sref{sec:background} key results of the modern theory of electric polarization, in \Sref{sec:review} we will consider possible mechanisms of inversion symmetry breaking in solids. The main conclusion is that the FM-FE state cannot be realized by magnetic means alone: the FM order is too simple to break $\mathcal{I}$. Then, in \Sref{sec:orbital}, we will turn to the orbital degrees of freedom and show how, by populating different combinations of orbitals at two sites of the bond, one can control not only the exchange coupling but also the electric polarization in the bond. Thus, orbital degrees of freedom appear to be that additional ingredient, which can help us in producing the FM-FE state. In \Sref{sec:practical}, we will turn to more practical aspects and consider the conditions, which should be met in order to realize the FM-FE state in solids: type of the lattice, electronic configuration of magnetic ions, type of the ligand atoms, details of electronic structure, etc. Particularly, the antiferro orbital order can be achived by minimizing the energy of exchange interactions~\cite{KugelKhomskii1972,KugelKhomskii1973,KugelKhomskii}, meaning that occupied orbitals should be flexible enough to adjust their shape. This flexibility can be created by intraatomic interactions responsible for Hund's second rule, which forces the atomic ground state to have maximal orbital degeneracy and competes with the Jahn-Teller (JT) distortion, acting in the opposite direction and tending to freeze occupied orbitals in one particular configuration. Then, in \Sref{sec:VI3}, we discuss results of numerical simulations for VI$_3$, using for these purposes realistic model derived from first-principle electronic structure calculations. Finally, in \Sref{sec:summary}, we summarize our material design strategy for ferromagnetic ferroelectrics and discuss possible implications of Hund's second rule for the properties of magnetic materials. 

\section{\label{sec:background} Theoretical background}
\par In quantum mechanics, the electric polarization is given by the expectation value of the position operator $\vec{r}$:
\noindent
\begin{displaymath}
\vec{P} = -\frac{e}{V} \langle \Psi  |\vec{r} \, | \Psi \rangle,
\end{displaymath}
\noindent where $-$$e$ is the electron charge, $V$ is the volume, and $\Psi$ is the ground state wavefunction. Since $\mathcal{I}$ transforms the polar vector $\vec{r}$ to $- \vec{r}$, the spontaneous polarization can develop only when $\mathcal{I}$ is macroscopically broken. In this sense, the search of new types of ferroelectric materials is essentially the search of new possibilities how to break $\mathcal{I}$. Some of such possibilities will be briefly reviewed in~\Sref{sec:review}.

\par According to the modern theory of electric polarization in periodic systems~\cite{FE_theory1,Resta}, $\vec{P}$ can be computed either in the $k$-space, via the Berry connection:
\noindent
\begin{equation}
\vec{P} = - \frac{ie}{(2 \pi)^3} \sum_{n = 1}^M
\int_{\rm BZ} \langle n \vec{k} | \vec{\nabla}_{\vec{k}} | n \vec{k} \rangle d \vec{k} ,
\label{eqn:PKSV}
\end{equation}
\noindent or, equivalently, in the $r$-space, via the Wannier functions $w_n$ for the occupied states:
\noindent
\begin{equation}
\vec{P} = - \frac{e}{V} \sum_{n = 1}^M
\int \vec{r} \, | w_n(\vec{r}\,) |^{2} d \vec{r}.
\label{eqn:PW}
\end{equation} 
\noindent The latter expression is important for understanding microscopic origin of magnetoelectric coupling. Since $\vec{r}$ does not depend on spin degrees of freedom, all information about magnetic state dependence of $\vec{P}$ is included in $w_n$. Equation~(\ref{eqn:PW}) is also useful for finding the analytical dependence of $\vec{P}$ on the magnetization in the framework of superexchange theory~\cite{PRB2020,PRL2021,MDPI2025}.

\section{\label{sec:review} Mechanisms of inversion symmetry breaking}
\subsection{Hybridization between bonding and antibonding states}
\par
\begin{figure}[t]
 \centering
\includegraphics[width=7cm]{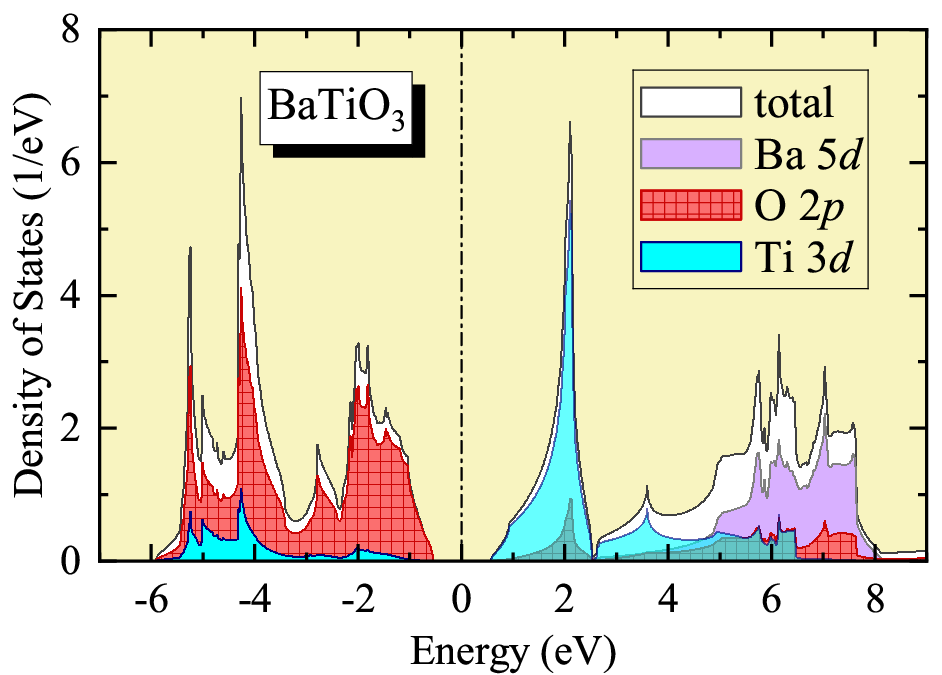} \includegraphics[width=7cm]{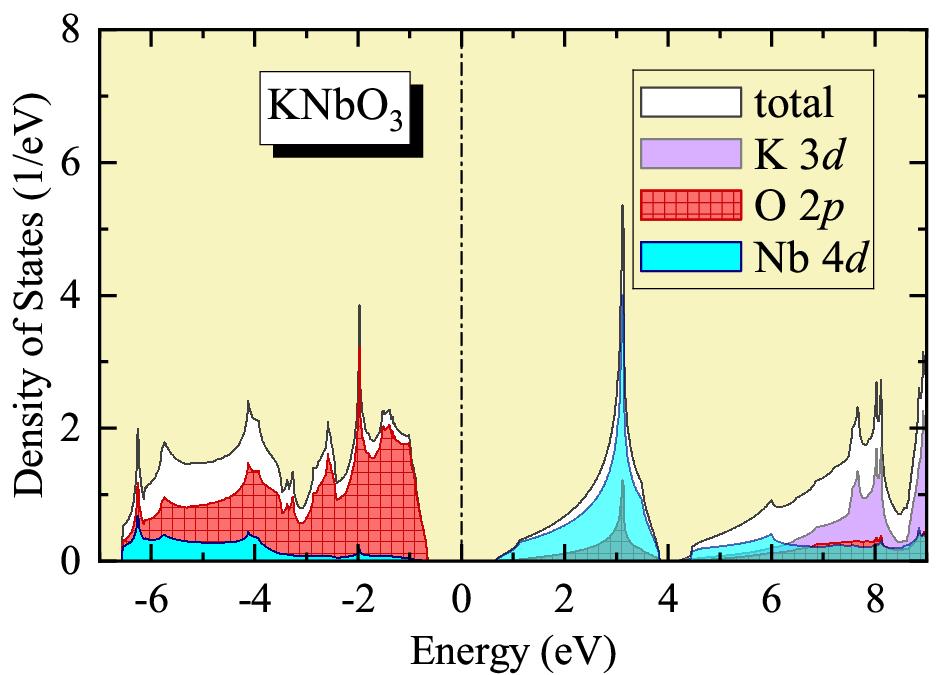} 
 \caption{Total and partial densities of states of cubic BaTiO$_3$ and KNbO$_3$ in the local density approximation. The Fermi level is in the middle of the band gap (shown by dot-dashed line).}
\label{fig:BaTiO3}
\end{figure}

 The first canonical example of FE materials is the so-called $d^0$ perovskites, which include such representative compounds as BaTiO$_3$ and KNbO$_3$~\cite{LinesGlass}. From the viewpoint of electronic structure in paraelectric cubic phase, the key aspect of these materials is the strong hybridization between the transition-metal $d$ and oxygen $p$ states, which splits them around the Fermi level, forming the bonding O $2p$ band and antibonding Ti $3d$ or Nb $4d$ band (\Fref{fig:BaTiO3}).  
\noindent
\noindent Because of this hybridization, both materials are insulators. Moreover, the bonding and antibonding states can be viewed as, respectively, symmetric and antisymmetric states relative to some inversion center. Then, the polar distortion, $\delta Q$, mixing these symmetric and antisymmetric states, will additionally shift the occupied O $2p$ band to the lower energy region, giving a possibility to realize the FE phase with spontaneously broken $\mathcal{I}$. This mechanism was proposed long ago by Bersuker using model considerations~\cite{Bersuker1966}. A transparent illustration was give by Cohen on the basis of first-principles electronic structure calculations~\cite{Cohen1992}. 

\par The corresponding energy gain is even in $\delta Q$. This is different from the conventional JT effect, which splits the degenerate states and, therefore, is odd in $\delta Q$~\cite{OpicPryce}. To emphasize this difference, the mechanism is called \emph{pseudo} JT effect. Then, the electronic energy gain should be combined with the energy loss, $\sim (\delta Q)^2$, resulting from the harmonic ion core motion~\cite{LinesGlass}. Depending on parameters, the total energy can have a minimum at finite $\delta Q$, signalling that the FE phase is stable and $\mathcal{I}$ is broken. 

\par However, the $d^0$ perovskites are intrinsically nonmagnetic. The magnetism would require a partial population of antibonding transition-metal band. Therefore, it will reduce the energy gain caused by mixing of bonding and antibonding bands, that will act against the instability towards the FE state. Nevertheless, certain populations of transition-metal states may still lead to the FE instability~\cite{Bersuker2012} and according to the first-principles calculations the ferromagnetic ferroelectricity can be indeed realized in SrMnO$_3$ under epitaxial strain~\cite{LeeRabe,EdstromEderer}.

\par Another important ingredient is the lone pair $6s^2$ electrons residing on the $A$ sites of such perovskites as PbTiO$_3$ and BiMnO$_3$~\cite{SeshadriHill}. The off-centrosymmetric displacements of $A$ sites induce the hybridization of the occupied $6s^2$ states with unoccupied states of opposite parity, which additionally stabilizes the polar phase.

\subsection{Type-II multiferroicity}
\par $\mathcal{I}$ can be broken by magnetic order. Such materials are called type-II multiferroics~\cite{Khomskii2009}. The canonical examples are spin-spiral insulators, where the onset of conical or cycloidal magnetic order gives rise to spontaneous polarization~\cite{Kimura_TbMnO3,TokuraSeki}. 

\par Thus, one can have a unique combination of ferroelectricity and magnetism, which can be used to mutually control each other applying either electric or magnetic field~\cite{TokuraScience}. However, literally, the multiferroicity means something different. It implies the combination of several \emph{ferroic} orders, so that the material should be not simply magnetic, but ferromagnetic. Therefore, the next question is: Can the FM order alone break the inversion symmetry?

\par
\begin{figure}[t]
 \centering
\includegraphics[width=14cm]{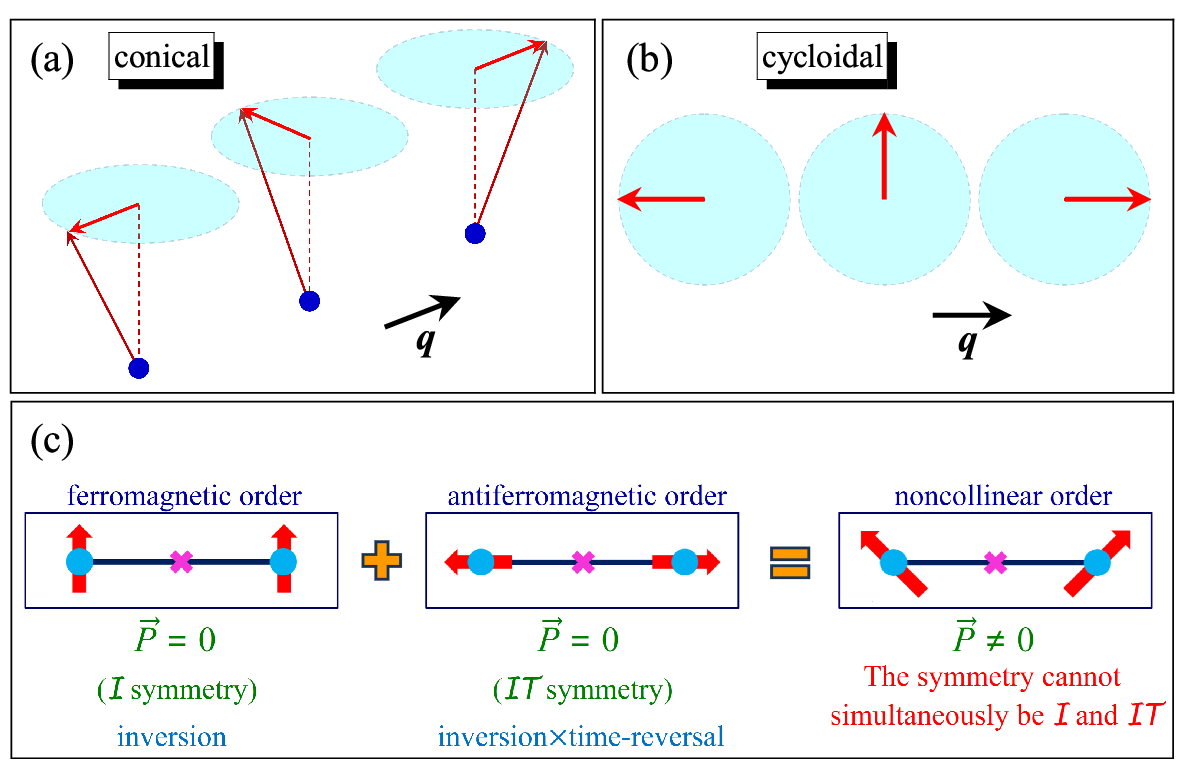} 
 \caption{Spin spirals and inversion symmetry breaking. (a) Conical and (b) cycloidal magnetic order. $\boldsymbol{q}$ is the propagation vector. (c) Illustration of inversion symmetry breaking by noncollinear spins in centrosymmetric bond. $\times$ is the inversion center. The noncollinear spin configuration can be decomposed in ferromagnetic and antiferromagnetic counterparts. The former is invariant under $\mathcal{I}$, while the latter is invariant under $\mathcal{IT}$. Since $\mathcal{I}$ cannot coexist with $\mathcal{IT}$, the inversion symmetry is broken (from~\cite{MDPI2025}).}
\label{fig:spiral}
\end{figure}

 To answer this question, one should understand why certain magnetic textures break $\mathcal{I}$. To be specific, let us consider the centrosymmetric bond connecting two noncollinear spins. This bond can be regarded as a part of cycloidal texture (\Fref{fig:spiral}). 
\noindent
\noindent Then, the noncollinear configuration of spins in the bonds can be presented as the superposition of collinear FM and AFM counterparts~\cite{MDPI2025}. While the FM configuration is transformed to itself by $\mathcal{I}$ about the bond center, the AFM configuration is transformed to itself by $\mathcal{IT}$, where $\mathcal{I}$ is combined with the time reversal $\mathcal{T}$, which additionally flips the spins.\footnote{Note $\mathcal{I}$ does not change the direction of spin, which is then axial vector.} However, $\mathcal{I}$ cannot coexist with $\mathcal{IT}$ because otherwise $\mathcal{T} \equiv \mathcal{IIT}$ would be another symmetry operation and the system would become nonmagnetic. The only possibility to resolve this conflict between FM and AFM counterparts coexisting in the noncollinear texture of spins is to break $\mathcal{I}$~\cite{MDPI2025}. This is the phenomenological reason why spiral magnetic order can induce the electric polarization. The microscopic explanation was proposed by Katsura, Nagaosa, and Balatsky~\cite{KNB}, which was later refined in the series of publications~\cite{PRL2021,MDPI2025,Xiang,KaplanMahanti}.

\par The rule is generic and applies not only to spin-spiral compounds but to all type-II multiferroics. In all these materials, the magnetic texture should involve simultaneously FM and AFM counterparts, transforming via, respectively, $\mathcal{I}$ and $\mathcal{IT}$~\cite{MDPI2025}. Therefore, the answer to the above question is \emph{no}: the FM order is just too simple to break the inversion symmetry.

\section{\label{sec:orbital} Orbital degrees of freedom}
\par Thus, the conventional type-II multiferroics cannot be ferromagnetic: in order to break $\mathcal{I}$ by spin degrees of freedom, it is necessary to destroy the ferromagnetism. This means that in order to achieve the FM-FE state, it is essential to consider additional degrees of freedom, which would control the magnetic coupling in the bonds \emph{and} simultaneously break $\mathcal{I}$. Below, we explore the potential of orbital degrees of freedom, which play a very important role in magnetism.

\subsection{Goodenough-Kanamori-Anderson rules and local breaking of inversion symmetry}
\par There are five $d$ orbitals and the magnetic properties of solids strongly depend on which orbitals are occupied and which are empty, and how the occupied orbitals are aligned relative to each other. The canonical example is the phenomenological GKA rules, proposed back in the 1950s~\cite{Anderson1950,Goodenough1955,Goodenough1958,Kanamori1959}. 

\par
\begin{figure}[t]
 \centering
\includegraphics[width=14cm]{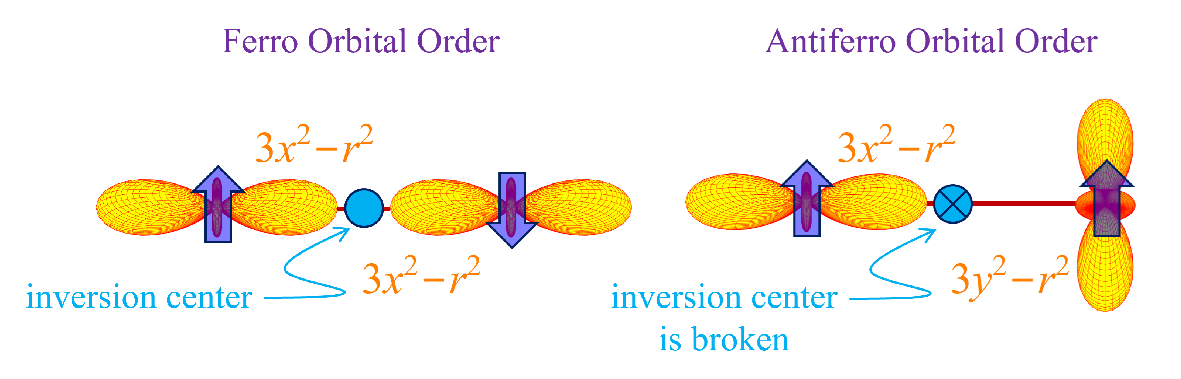} 
 \caption{Examples of ferro and antiferro orbital order around the inversion center. Ferro orbital order tends to stabilize AFM coupling and preserve the inversion symmetry in the bond. Antiferro orbital order stabilizes FM coupling and breaks the inversion symmetry.}
\label{fig:OO}
\end{figure}

 These rules state essentially that if electrons occupy the same orbitals at two sites of the bond (the so-called ferro orbital order), the exchange coupling will be antiferromagnetic (\Fref{fig:OO}). If they occupy different orbitals, aligned perpendicular to each other (antiferro orbital order), the exchange coupling will most likely be ferromagnetic.  
\noindent

\par However, there is one important point, which was overlooked in this canonical GKA picture: the antiferro orbital order not only stabilizes the FM coupling but, since the occupied orbitals across the inversion center become inequivalent, also breaks $\mathcal{I}$ in the bond. Thus, at least in the single bond, one can easily realize the FM-FE state.

\subsection{Superexchange theory}
\par The superexchange theory provides the microscopic explanation for the GKA rules~\cite{Anderson1959}. It starts with the atomic limit, where occupied and empty states are splits by the on-site Coulomb repulsion $U$ and intraatomic exchange interaction $J_{\rm H}$, and treats transfer integrals between them as a perturbation. $U$ in such a picture enforces the charge neutrality of atomic configuration with the integer number of electrons, while $J_{\rm H}$ is responsible for Hund's first rule~\cite{PRB94}. 

\par
\begin{figure}[t]
 \centering
\includegraphics[width=14cm]{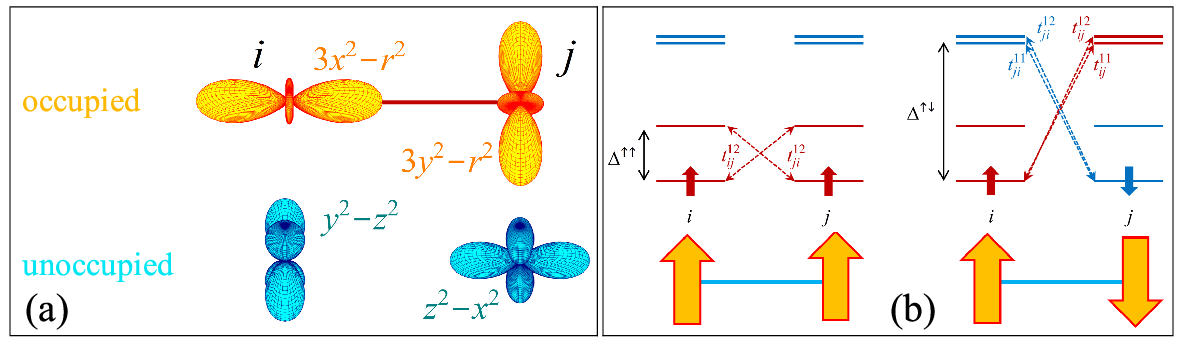} 
 \caption{Illustration of superexchange interactions: (a) Example of occupied and unoccupied orbitals in the bond; (b) Transfer integrals between occupied and empty states in the case of FM and AFM alignment. The splitting is $\Delta^{\uparrow \uparrow} = U - J_{\rm H}/2 \equiv \tilde{U}$ and $\Delta^{\uparrow \downarrow} = U + J_{\rm H}/2$.}
\label{fig:SE}
\end{figure}

 Then, the exchange coupling $J_{ij}$ is obtained in the second-order perturbation theory for the energy difference between FM and AFM configurations in the bond. For instance, considering the superexchange processes involving the occupied $3x^2-r^2$ ($3y^2-r^2$) and unoccupied $y^2-z^2$  ($z^2-x^2$) $e_{g}$ orbitals at the site $i$ ($j$) (see \Fref{fig:SE}), $J_{ij}$ will be given by
\noindent
\begin{displaymath}
J_{ij} = \frac{x [ (t_{ij}^{12})^{2}+(t_{ji}^{12})^{2} ]- 2(t_{ij}^{11})^{2}}{2\tilde{U}(1+x)},
\end{displaymath}
\noindent where $t_{ij}^{11} = -1/2$, $t_{ij}^{12} = -\sqrt{3}/2$, and $t_{ji}^{12} = 0$ are the transfer integrals between occupied ($1$) and unoccupied ($2$) orbitals in terms of the two-center Slater-Koster integral $dd \sigma$~\cite{SlaterKoster}, $\tilde{U} = U - J_{\rm H}/2$, and $x = J_{\rm H}/\tilde{U}$.\footnote{The corresponding exchange energy in the bond is defined as $-J_{ij} \boldsymbol{e}_{i} \cdot \boldsymbol{e}_{i}$, where $\boldsymbol{e}_{i}$ is the unit vector in the direction of spin magnetic moment at the site $i$. Furthermore, following previously adopted notations~\cite{MDPI2025}, we use vector symbols, such as $\vec{P}$, to denote polar vectors and bold symbols, such as $\boldsymbol{e}$, to denote pseudovectors.}
\noindent
\noindent Thus, the coupling is ferromagnetic when $2 J_{\rm H} > U$. The tendency towards ferromagnetism is further enhanced if there are other $d$ electrons, which tend to form high-spin state (due to $J_{\rm H}$) and, therefore, favor the FM configuration when electrons transfer to neighboring sites with the same direction of spin. For instance, when beside $e_{g}$ there are inner $t_{2g}$ electrons, which interact via $J_{\rm H}$, but do not participate in the hoppings processes, $J_{ij}$ is described by the same equation but with $x = NJ_{\rm H}/\tilde{U}$, where $N$ is the total number of $d$ electrons. Then, the FM coupling is stabilized when $\frac{3N+1}{2}J_{\rm H} > U$. Such a situation is realized in perovskite manganites~\cite{JKPS1998}. 

\par These are typical considerations for exchange interactions, which were known for decades. Nevertheless, similar arguments can be applied also for $\vec{P}$, starting from the expression (\ref{eqn:PW}) of the modern theory of electric polarization~\cite{FE_theory1,Resta}. Indeed, in the first-order perturbation theory, the occupied Wannier function at the site $i$ will be given by
\noindent
\begin{displaymath}
| w_{i} \rangle = | \alpha_{i}^{1} \rangle + \sum\limits_{j \ne i} | \alpha_{i \to j}^{1} \rangle ,
\end{displaymath}
\noindent where $\alpha_{i}^{1}$ is the ``head'' residing at the central site $i$ and $\alpha_{i \to j}^{1}$ are the ``tails'' induced by the electron hoppings at the neighboring sites $j$. The latter can be evaluated using the perturbation theory as 
\noindent
\begin{displaymath}
| \alpha_{i \to j}^{1} \rangle = -\frac{t_{ij}^{12}}{\tilde{U}} | \alpha_{j}^{2} \rangle.
\end{displaymath} 
\noindent Then, assuming $\langle \alpha_{i}^{a}| \vec{r} \, | \alpha_{j}^{b} \rangle \approx \vec{R}_{j} \delta_{ij} \delta_{ab}$~\cite{PRB2020} and using \Eref{eqn:PW}, it is straightforward to find the following expression for electric polarization in the bond $ij$:
\noindent
\begin{equation}
\vec{P}_{ij}=\frac{e}{V}\frac{(t_{ij}^{12})^2-(t_{ji}^{12})^2}{\tilde{U}^{2}} (\vec{R}_{i}-\vec{R}_{j}).
\end{equation}
\noindent Thus, the polarization is finite if $t_{ij}^{12} \ne t_{ji}^{12}$, i.e. when the hoppings between occupied and empty orbitals are nonreciprocal. Such nonreciprocity is produced by the antiferro orbital order, which makes the directions $i \to j$ and $j \to i$ inequivalent.

\subsection{Kugel-Khomskii theory}
\par Kugel-Khomskii theory is basically a generalization of the superexchange theory, treating spin and orbital degrees of freedom on an equal footing~\cite{KugelKhomskii1972,KugelKhomskii1973,KugelKhomskii}. If orbital degeneracy is lifted by the lattice distortion, deciding which orbitals are occupied and which are empty, it is reduced to the conventional superexchange theory, specifying the exchange coupling for the given configuration of occupied orbitals. In this case, the lattice distortion acts as an external constraining field in the spin-orbital system. However, if the lattice distortion is weak, there exists certain self-organization mechanism, when the system decides by itself the type of the relative alignment of the occupied orbitals, by minimizing the energy of exchange interactions with respect to the orbital degrees of freedom for the given configuration of spins. If this orbital alignment appears to be of the antiferro type, it can stabilize the FM coupling and induce the electric polarization as described above. 

\section{\label{sec:practical} Towards practical realization}
\par In the previous Section we have seen that the FM-FE state can be easily realized in the single bond by the antiferro orbital order. In this Section, we will discuss whether such state can be realized in real materials having periodic structure. 

\subsection{Crystal lattice: perovskite versus honeycomb}
\par
\begin{figure}[t]
 \centering
\includegraphics[width=14cm]{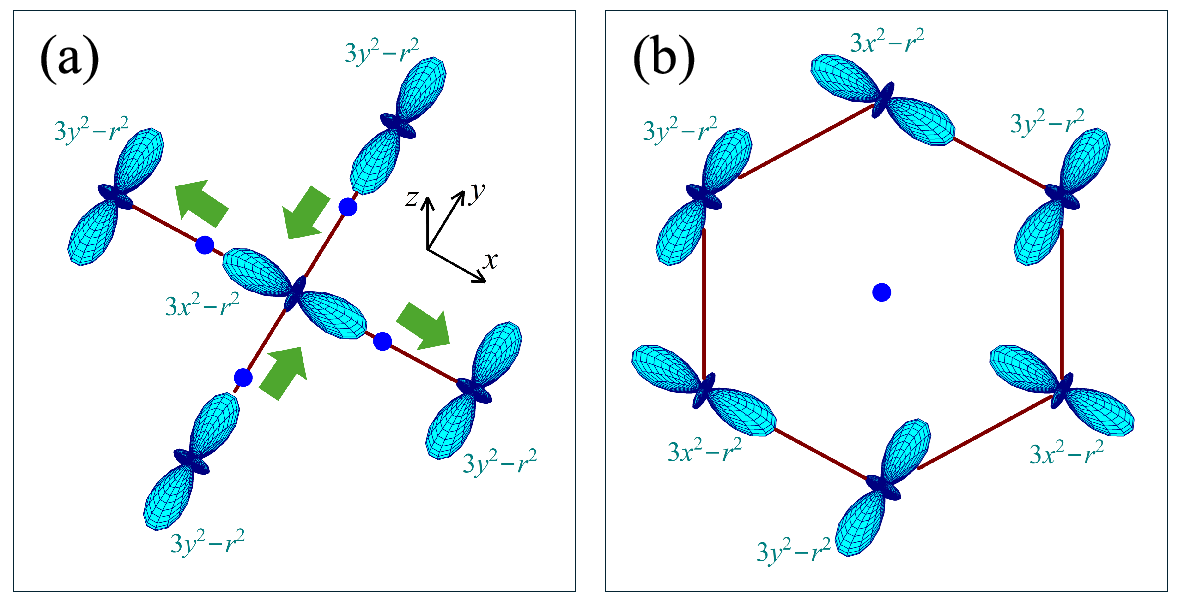} 
 \caption{Example of the antiferro orbital order realized in perovskite (a) and honeycomb (b) lattice. In the perovskite lattice, the directions of polarization induced in each bond are shown by arrows. The inversion centers are shown by circles (which coincide with the positions of oxygen atoms in undistorted perovskite structure).}
\label{fig:Lattices}
\end{figure}

 The orbital order has been intensively studied in magnetoresistive manganites and other transition-metal compounds crystallizing in the perovskite structure~\cite{KhomskiiStreltsov,PRL96,Maezono1998,Mochizuki2003,Sawada1996,Blake2001,review2008}. Particularly, the antiferro orbital order is believed to be responsible for the FM character of exchange interactions in LaMnO$_3$~\cite{Goodenough1955,PRL96}, YTiO$_3$~\cite{Mochizuki2003}, LaVO$_3$~\cite{Sawada1996}, and YVO$_3$~\cite{Blake2001}. However, in all these materials the magnetic atoms are located in the inversion centers. Therefore, although the electric polarization can be induced in each bond, there will always be another bond with the opposite direction of electric polarization, as explained in~\Fref{fig:Lattices}(a).   
\noindent
\noindent Thus, most likely, these perovskite materials will be \emph{antiferroelectric}. They can host very interesting properties, such as weak ferromagnetism, net orbital magnetization, etc.~\cite{arXiv2025}, but hardly suitable from the viewpoint of FE applications. 

\par The honeycomb lattice seems to be more promising. In this case, $\mathcal{I}$ connects two magnetic sublattices, while the atoms themselves are located not in the inversion centers, as explained in~\Fref{fig:Lattices}(b). Therefore, if we succeed in realizing the antiferro orbital order between the sublattices, our system can become simultaneously ferromagnetic and ferroelectric.  

\subsection{Electronic configuration: $d^{1}$ versus $d^{2}$}
\par
\begin{figure}[t]
 \centering
\includegraphics[width=14cm]{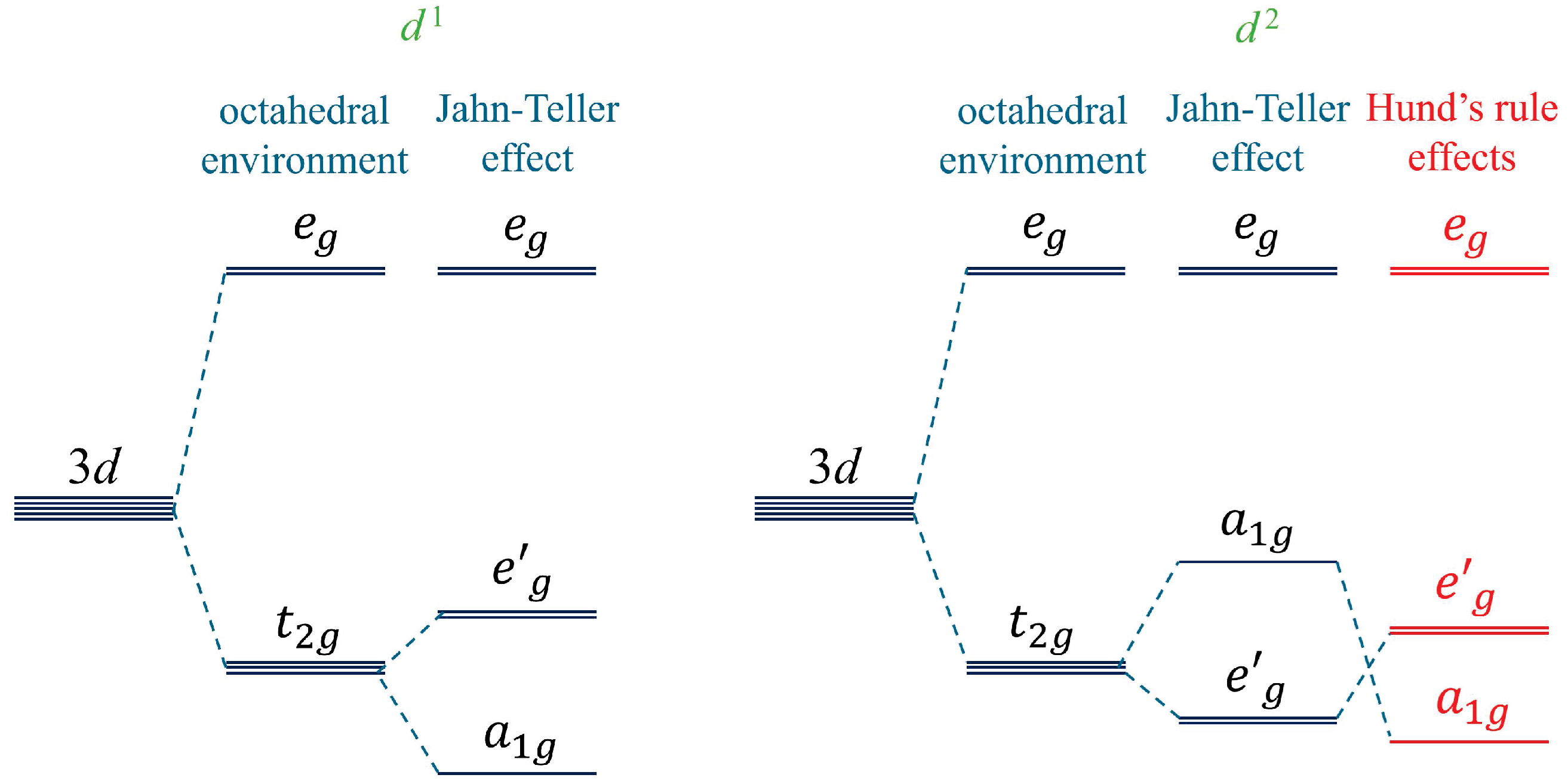} 
 \caption{Atomic level splitting for electronic configurations $d^{1}$ (left) and $d^{2}$ (right). The octahedral environment of ligands around the transition-metal sites will split the $3d$ levels into threefold degenerate $t_{2g}$ and twofold degenerate $e_{g}$ states, correspondingly in the lower and upper part of the spectrum (the so-called $10Dq$ splitting). The Jahn-Teller distortion tends to lift the orbital degeneracy and further split the $t_{2g}$ manifold into twofold degenerate $e'_{g}$ and nondegenerate $a_{1g}$ states ($\Delta_{\rm tr}$ splitting). This distortion works in the opposite directions for $d^{1}$ and $d^{2}$, stabilizing, respectively, $a_{1g}$ and $e'_{g}$ states. The $d^{2}$ configuration is subjected to atomic Hund's second rule effects, which tend to reenforce the ground state with maximal multiplicity and reverse the order of $a_{1g}$ and $e'_{g}$ states.}
\label{fig:atconfiguration}
\end{figure}

 The next important question is how to realize the antiferro orbital order on the honeycomb lattice. First, we explore the conventional picture based on the JT distortion, which takes place, for instance, in LaMnO$_3$~\cite{Kanamori1960}. We assume that the transition-metal atoms are in the octahedral environment, which results in the $10Dq$ splitting of $3d$ levels into threefold degenerate $t_{2g}$ and twofold degenerate $e_{g}$ states, as explained in~\Fref{fig:atconfiguration}.
\noindent
\noindent Then, we consider the situation when there is either one or two $3d$ electrons residing in the $t_{2g}$ shell, corresponding to the electronic configurations $d^{1}$ and $d^{2}$. Thus, the system is subjected to the JT distortion, which further splits degenerate $t_{2g}$ levels into twofold degenerate $e'_{g}$ and nondegenerate $a_{1g}$ states. The corresponding parameters is denoted as $\Delta_{\rm tr}$. Moreover, the JT theorem states in this respect that the distorted system should have a nondegenerate ground state~\cite{JT}. Therefore, the JT distortion should work in opposite directions for the configurations $d^{1}$ and $d^{2}$, stabilizing, respectively, $a_{1g}$ and $e'_{g}$ states. Another important point is that for two sites connected by $\mathcal{I}$ the JT distortion is expected to be the same. Thus, the JT mechanism can stabilize ferro orbital order but not the antiferro one and alone does not break the inversion symmetry. Furthermore, since the ground state is nondegenerate, the orbital degrees of freedom are frozen. This suppress the Kugel-Khomskii mechanism of the orbital ordering when the latter is driven by superexchange processes.
 
\par Fortunately, the JT distortion is relatively weak in $t_{2g}$ systems so that other mechanisms can easily compete with it. One of such prominent mechanisms is Hund's second rule, driven by intraatomic interactions. In isolated atoms, the proper interaction parameter (the Racah parameter) is given by $B = \frac{9F^{2}-5F^{4}}{441}$, in terms of the radial Slater integrals $F^{2}$ and $F^{4}$~\cite{Racah,Slater}. For comparison, the intraatomic exchange coupling responsible for Hund's first rule is $J_{\rm H} = \frac{F^{2}+F^{4}}{14}$. Since $F^{4} \sim 0.63 F^{2}$~\cite{PRB94}, the ratio $\frac{B}{J_{\rm H}}$ is about $0.1$, which naturally explains the hierarchy of atomic Hund's rule, where the second rule should be considered only after the first one. Since for $3d$ ions $J_{\rm H} \sim 1$ eV~\cite{PRB94,KhomskiiStreltsov}, $B$ is about $0.1$ eV. The important point is that typical crystal-field splitting of $t_{2g}$ levels caused by the JT distortion is also of the order of $0.1$ eV~\cite{review2008}. Therefore, these two mechanisms can compete with each other. 

\par While the JT distortion tends to quench the orbital degrees in one particular configuration, Hund's second rule acts in the opposite direction and tends to realize the ground state with maximal orbital degeneracy. In terms of the one-electron picture in~\Fref{fig:atconfiguration}, it can revert the order of the $a_{1g}$ and $e'_{g}$ levels for the configuration $d^{2}$. Therefore, in the atomic limit, one electron will reside on the nondegenerate $a_{1g}$ orbital, while another one will be in the degenerate subspace spanned by two $e'_{g}$ orbitals.\footnote{A rigorous picture can be obtained by considering exact two-electron states~\cite{PRB2024}, following the arguments of Tanabe and Sugano~\cite{TanabeSugano}.} This will reactivate the Kugel-Khomskii mechanism of the orbital ordering so that the occupied orbitals can adjust their shape to minimize the energy of exchange interactions and, hopefully, realize the antiferro order, breaking the inversion symmetry. Simple toy-model considerations supporting this idea can be found in~\cite{PRB2024}.

\par The Hund's rules are essentially many-electron effects. They do not operate in the one-electron configuration $d^{1}$, where the JT distortion is the only mechanism that splits the $t_{2g}$ levels. Therefore, in order to realize the antiferro orbital order on the honeycomb lattice, it is important to consider two-electron (or other) systems, in which Hund's rule effects are operative. 

\subsection{Details of electronic structure: V$_2$O$_3$ versus VI$_3$}
\par In this Section we will consider two potential $d^{2}$ candidates, V$_2$O$_3$ and VI$_3$, and argue that the antiferro orbital order stabilizing the FM-FE state can be probably realized in VI$_3$ but not in V$_2$O$_3$. The reason lies in details of the electronic structure, which is also related to the differences in the crystal structure and types of the ligand atoms. 

\par At room temperature, V$_2$O$_3$ crystallizes in corundum $R\overline{3}c$ structure~\cite{V2O3} while VI$_3$ is the van der Waals crystal having the space group $R\overline{3}$~\cite{VI3}. Thus, the main structural motif of VI$_3$ is honeycomb layers where two V$^{3+}$ sublattices are connected by $\mathcal{I}$. The ions in the same sublattices are connected by threefold rotations and translations. The corundum structure cannot be presented as a simple combination of the honeycomb planes. There are two pairs of the V$^{3+}$ sublattices. The ions in each pair are connected by $\mathcal{I}$. Each sublattice is transformed to itself by threefold rotations and translations, like in VI$_3$. However, the stacking and coordination is different from the regular honeycomb materials~\cite{Mattheiss}. 

\par Using electronic structure in the local density approximation (LDA), one can construct the effective model for the V $3d$ bands in V$_2$O$_3$ and VI$_3$, which are primarily responsible for the magnetism. Namely, the one-electron parameters can be related to the matrix elements of LDA Hamiltonian in the Wannier basis for the V $3d$ bands located in the interval $[-1,4]$ eV for V$_2$O$_3$ and $[-0.5,2]$ eV for VI$_3$. The effective Coulomb interactions can be evaluated using constrained random-phase approximation (cRPA)~\cite{cRPA}. Technical details can be found in~\cite{review2008}. 

\par
\begin{figure}[t]
 \centering
\includegraphics[width=7cm]{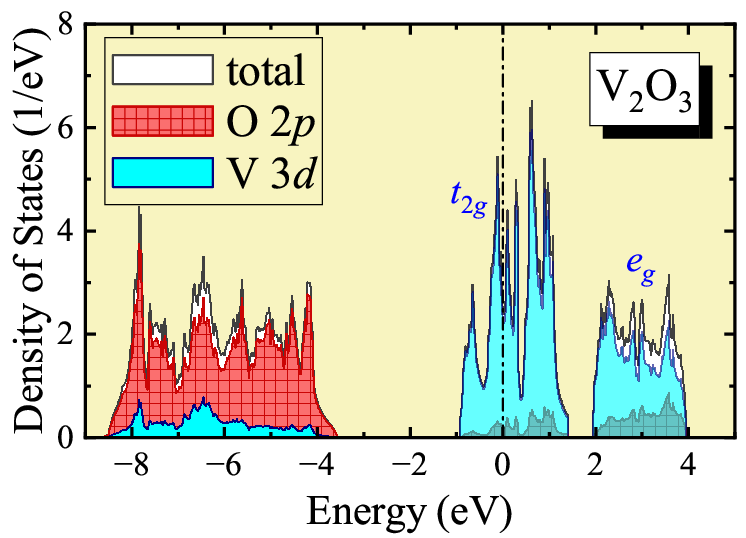} \includegraphics[width=7cm]{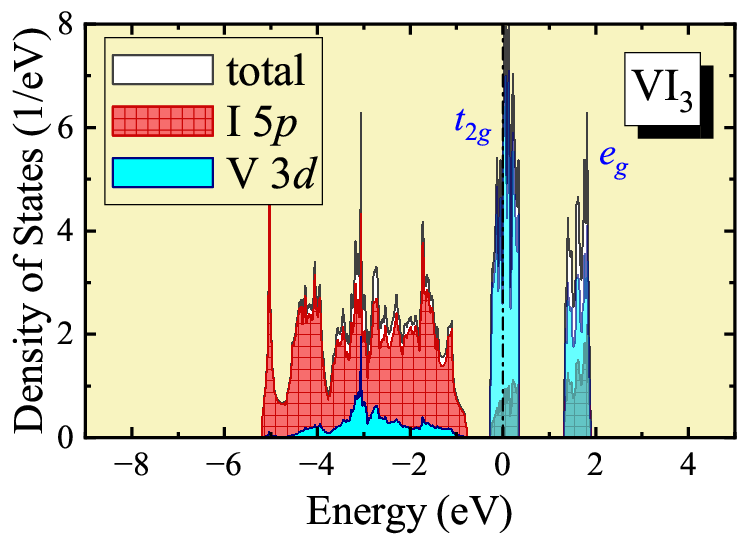} 
 \caption{Total and partial densities of states of V$_2$O$_3$ and VI$_3$ in the local density approximation. The Fermi level is at zero energy (shown by dot-dashed line).}
\label{fig:VI3DOS}
\end{figure}

 The model parameters depend on the electronic structure, which appears to be very different in V$_2$O$_3$ and VI$_3$ (\Fref{fig:VI3DOS}).
\noindent
\noindent Namely: 
\begin{itemize}
\item The interaction parameters for V $3d$ bands in cRPA depend on how well they are screened by other bands, particularly by O $2p$ and I $5p$ bands. The closer the bands are, the stronger the screening. Thus, the parameters $U$, $J_{\rm H}$, and $B$ are generally weaker in VI$_3$ (table~\ref{tab:H}). Moreover, $J_{\rm H}$ and $B$ are typically less screened than $U$. Therefore, the ratio $\frac{J_{\rm H}}{U}$ is substantially larger in VI$_3$, which is important for stabilizing FM interactions~\cite{KugelKhomskii};
\item The trigonal splitting between $e'_{g}$ and $a_{1g}$ states ($\Delta_{\rm tr}$) is substantially smaller in VI$_3$, so that $\frac{\Delta_{\rm tr}}{B} < 1$ in VI$_3$ but $> 1$ in V$_2$O$_3$. This will reactivate the Kugel-Khomskii mechanism and drive the formation of antiferro orbital order in VI$_3$ but not in V$_2$O$_3$;
\item $10Dq$ splitting is also smaller in VI$_3$. This facilitates the mixing of the $e'_{g}$ and $e_{g}$ states belonging to the same representation by Hund's rule interactions and further minimizes the energy of these interactions. This further support the Kugel-Khomskii mechanism and formation of antiferro orbital order in VI$_3$.
\end{itemize}
\noindent
\begin{table}[t]
\caption{\label{tab:H} Parameters of model Hamiltonian for V$_2$O$_3$ and VI$_3$ (in eV): trigonal splitting between $e'_{g}$ and $a_{1g}$ states ($\Delta_{\rm tr}$), $10Dq$ splitting between $t_{2g}$ and $e_{g}$ states, on-site Coulomb repulsion $U$, intraatomic exchange coupling $J_{\rm H}$ responsible for Hund's first rule, and Racah parameter $B$ responsible for Hund's second rule.}
\begin{indented}
\lineup
\item[]\begin{tabular}{@{}cccccc}
\br
compound   & $\Delta_{\rm tr}$ & $10Dq$ & $U$    & $J_{\rm H}$ & $B$     \cr
\mr
V$_2$O$_3$ & $0.15$            & $2.31$ & $3.05$ & $0.85$      & $0.09$  \cr
VI$_3$     & $0.01$            & $1.49$ & $1.21$ & $0.75$      & $0.07$  \cr
\br
\end{tabular}
\end{indented}
\end{table}

\par Thus, VI$_3$ appears to be a good candidate for spontaneous breaking of inversion symmetry by the antiferro orbital order and realizing the FM-FE state. The $d^{2}$ configuration itself does not necessarily guarantee the emergence of  antiferro orbital order, where many things depend on details of the electronic structure.

\section{\label{sec:VI3} What can one expect from VI$_3$?}
\par Finally, we turn to numerical mean-field Hartree-Fock (HF) calculations for the model, which was obtained for the V $3d$ bands of VI$_3$ as described above. Some of the model parameters are summarized in table~\ref{tab:H}. The total energy in the HF approximation is expressed in terms of the one-electron density matrix at each site of the lattice, $\hat{n} = [n_{ab}^{\sigma \sigma'}]$, which is calculated self-consistently, where $a$ and $b$ are the orbital indices (two $e_{g}$, one $a_{1g}$, and two $e'_{g}$), while $\sigma$ and $\sigma'$ are the spin indices ($\uparrow$ or $\downarrow$). The nondiagonal matrix elements with respect to $\uparrow$ and $\downarrow$ are typically induced by relativistic spin-orbit (SO) interaction. Otherwise, $\hat{n}$ is diagonal. The details can be found in~\cite{review2008}.

\par
\begin{figure}[t]
 \centering
\includegraphics[width=14cm]{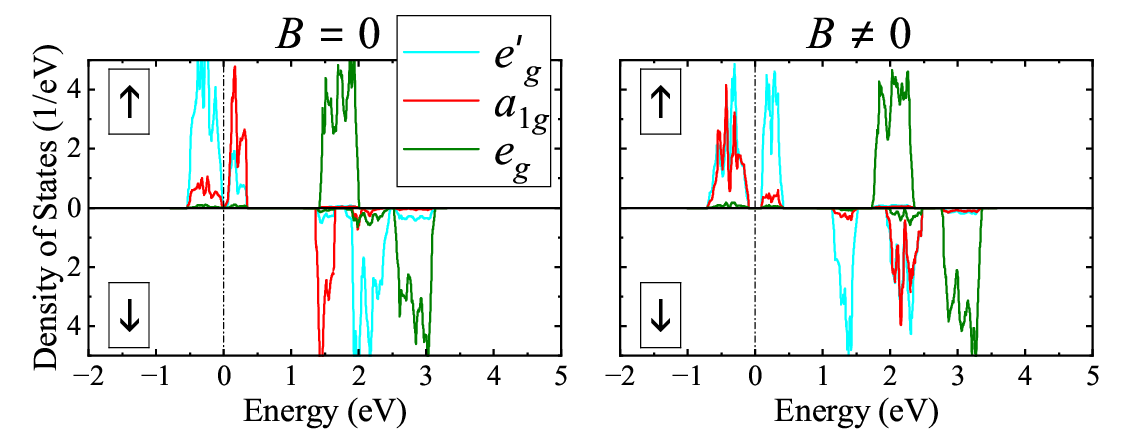} 
 \caption{Total and partial densities of states as obtained in the Hartree-Fock approximation for the ferromagnetic state, where the Racah parameter $B$ responsible for Hund's second rule was set to either $0$ (left) or $0.07$ eV (right). The Fermi level is at zero energy (the middle of the band gap).}
\label{fig:HF}
\end{figure}

 The corresponding densities of states for the FM state with and without the on-site interactions responsible for Hund's second rule are shown in~\Fref{fig:HF}.
\noindent
\noindent The strength of these interactions is controlled by the Racah parameter $B$. As expected, when $B=0$, the type of occupied states is solely determined by weak trigonal splitting $\Delta_{\rm tr}$, so that two $3d$ electrons tend to occupied twofold degenerate $e'_{g}$ states. Therefore, the ground state appears to be nondegenerate. The $a_{1g}$ states are located mainly in the unoccupied part and mixed with $e'_{g}$ ones by intersite hoppings. However, the situation changes dramatically when $B$ is finite, as schematically explained in~\Fref{fig:atconfiguration}. In this case, two $3d$ electrons reside on the $a_{1g}$ orbital and one of the $e'_{g}$ orbital. Thus, the ground state is degenerate. The shape of the occupied $e'_{g}$ states is controlled by superexchange interactions, which further lower the energy via the antiferro orbital ordering~\cite{KugelKhomskii1972,KugelKhomskii1973,KugelKhomskii}. 

\par Similar picture can be obtained using more sophisticated dynamical mean-field theory (DMFT)~\cite{PRB2024}.

\subsection{Orbital ordering and inversion symmetry breaking}
\par
\begin{figure}[t]
 \centering
\includegraphics[width=14cm]{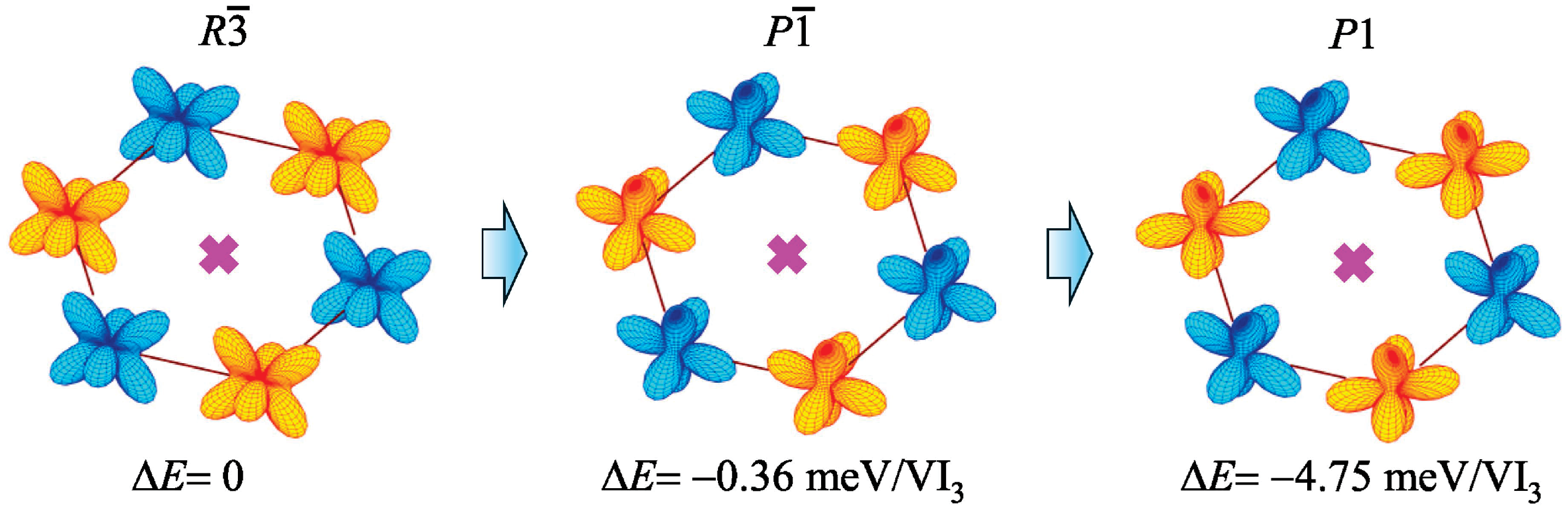} 
 \caption{Orbital ordering in the ferromagnetic state of VI$_3$ as obtained in model Hartree-Fock calculations by enforcing the original trigonal $R\overline{3}$ symmetry, the triclinic $P\overline{1}$ symmetry, and fully relaxing the symmetry ($P1$). The crystallographic inversion centers are denoted by $\times$. To sublattices of honeycomb lattice are displayed by different colors. $\Delta E$ is the corresponding energy change relative to the state with the $R\overline{3}$ symmetry (from~\cite{PRB2024}).}
\label{fig:HFOO}
\end{figure}

 Let us start with restricted HF calculations without SO interaction, where we additionally constrain the form of $\hat{n}$ to satisfy the $R\overline{3}$ symmetry of VI$_3$ lattice. Technically, $\hat{n}$ is averaged by the matrices of threefold rotations in each sublattices to guarantee the $R3$ symmetry. Then, $\hat{n}$ is averaged over the sublattices to guarantee the inversional symmetry. This $R\overline{3}$ solution is regarded as the reference point (\Fref{fig:HFOO}). 
\noindent

\par Then, we relax the constraints. First, we turn off the rotational constraint and average $\hat{n}$ only over the sublattices, enforcing the inversional invariance. The atoms within the sublattices are connected by translations, but no longer by the threefold rotations. The corresponding symmetry of the orbital order is $P\overline{1}$. It changes the shape of the occupied orbitals, but only slightly lowers the energy.

\par Finally, we relax the inversional constraint and treat $\hat{n}$ in two sublattices as independent variables. The corresponding symmetry is $P1$. In comparison with the $P\overline{1}$ state, the occupied orbitals in two sublattices are additionally rotated relative to each other, so that $\mathcal{I}$ becomes broken, resulting in significant energy gain (\Fref{fig:HFOO}).

\par
\begin{figure}[t]
\begin{center}
\includegraphics[width=14cm]{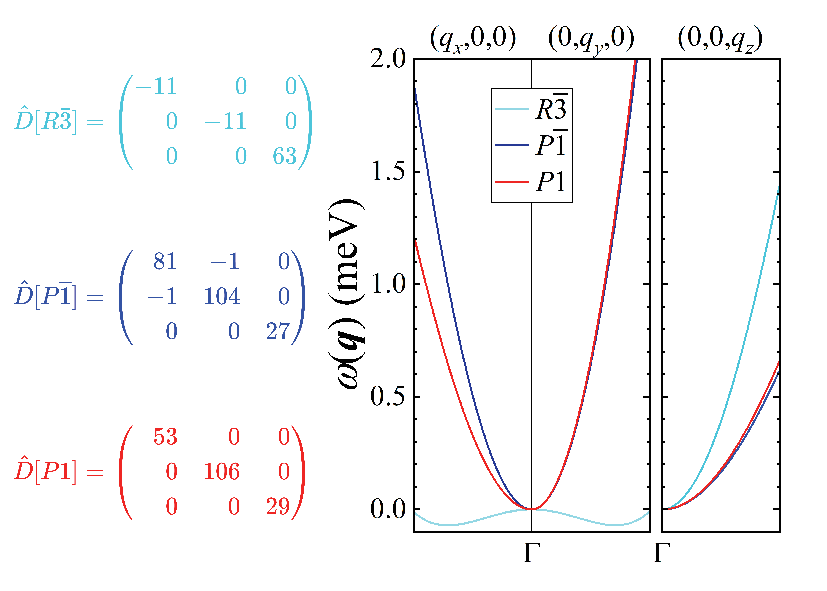} 
\end{center}
\caption{
Spin-wave stiffness tensors (in meV\AA$^2$) for the orbital states of the $R\overline{3}$, $P\overline{1}$, and $P1$ symmetry, and corresponding spin-wave dispersions near the $\Gamma$-point of Brillouin zone (from~\cite{PRB2024}).}
\label{fig.sw}
\end{figure}

 The next important question is whether the FM spin order is stable or not. The answer depends on the spin-wave stiffness $\hat{D}$, which can be evaluated using linear response theory for interatomic exchange interaction~\cite{review2024}. In the present context, it is basically a perturbation theory with respect to infinitesimal rotations of spins. If $\hat{D}$ is positive definite, the FM order is stable. If not, it is unstable. The results are summarized in~\Fref{fig.sw}.
\noindent
\noindent One can clearly see that the FM state is unstable if the orbital configuration respects the $R\overline{3}$ symmetry of the lattice. Nevertheless, lowering the orbital symmetry will stabilize the FM state.  

\subsection{Magnetic field control of electric polarization}
\par In the previous Section we have seen that, in the FM state of VI$_3$, $\mathcal{I}$ can be broken by the orbital order. The next important questions are: (i) How large is the electric polarization? (ii) How can it be controlled by a magnetic field?

\par
\begin{figure}[t]
\begin{center}
\includegraphics[width=14cm]{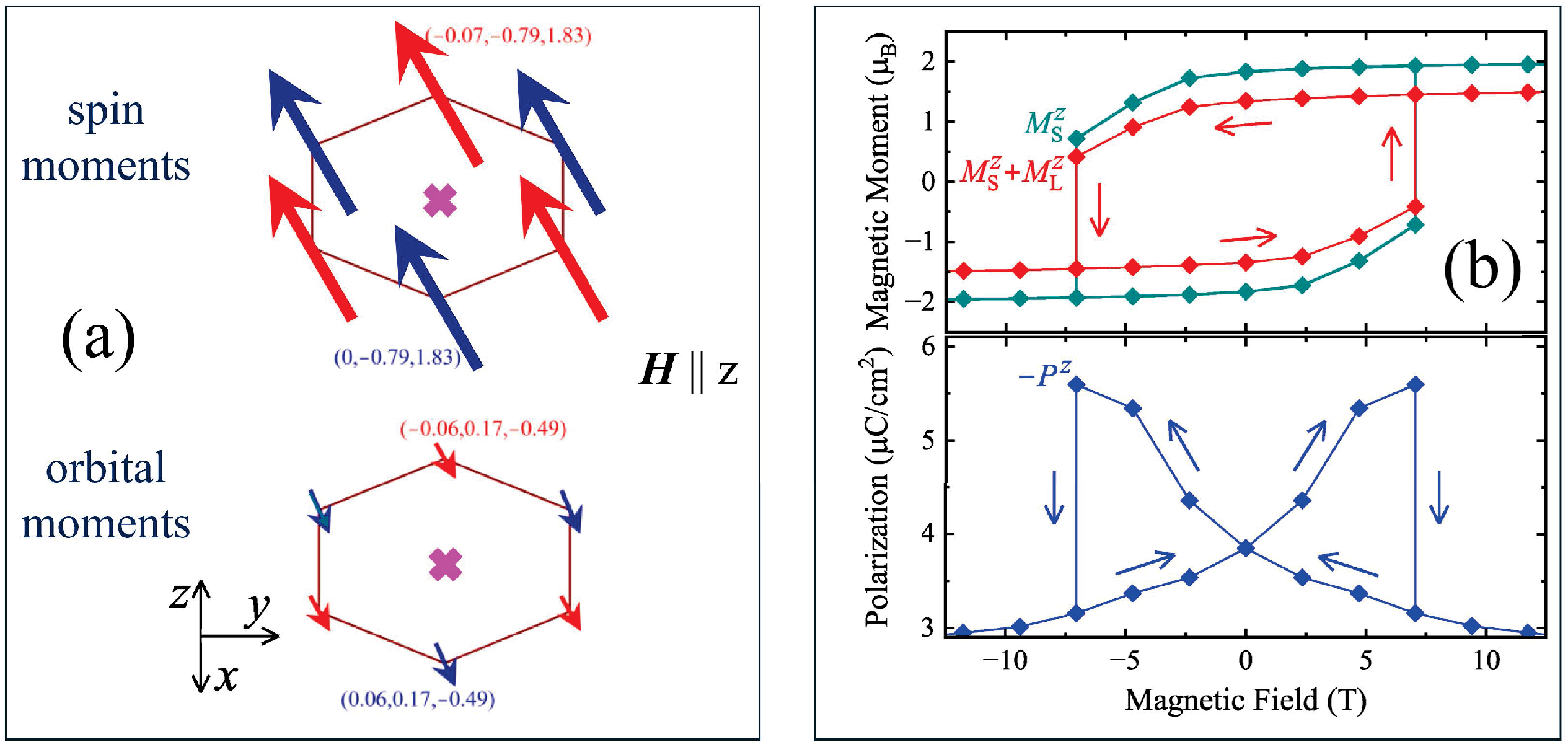} 
\end{center}
\caption{Results of Hartree-Fock simulations with spin-orbit interaction and magnetic field. (a) Directions of spin ($\boldsymbol{M}_{\rm S}$) and orbital ($\boldsymbol{M}_{\rm L}$) magnetic moments without magnetic field. Two sublattices are denoted by red and blue colors. The numerical values of $\boldsymbol{M}_{\rm S}$ and $\boldsymbol{M}_{\rm L}$ are given in  parentheses. The crystallographic inversion center is denoted by $\times$. (b) Magnetic-field dependence of magnetization (top) and electric polarization (bottom). The field is applied parallel to $z$. $M^{z}_{\rm S}$, $M^{z}_{\rm S}$$+$$M^{z}_{\rm L}$, and $P^{z}$ are the values of, respectively, spin magnetic moment, total magnetic moment, and electric polarization along $z$ (from~\cite{PRB2024}).}
\label{fig.H}
\end{figure}

 The key point here is that the orbital order breaks not only inversional but also the threefold rotational symmetry. Therefore, the magnetic moments are not necessarily aligned perpendicular to the honeycomb plane or lie in this plane (\Fref{fig.H}). 
\noindent
\noindent Since the rotational symmetry is broken by the orbital order, the $a_{1g}$ and $e'_{g}$ states do not longer belong to different representations and are allowed to mix. The angle between magnetic moments and the hexagonal $z$-axis is induced by the SO interaction and depends on the degree of this mixing. This angle can be also controlled by the external magnetic field $\boldsymbol{H} = (0,0,H)$ applied parallel to $z$. Therefore, $\boldsymbol{H}$ can also control the electronic degrees of freedom, such as the $a_{1g}$-$e'_{g}$ mixing and the polarization $\vec{P}$. This is the main mechanism of magneto-electric coupling in the symmetry-broken state of VI$_3$. Furthermore, besides spin magnetic moment, $|\boldsymbol{M}_{\rm S}| \sim 2~\mu_{\rm B}$, there is an appreciable orbital magnetic moment, $|\boldsymbol{M}_{\rm L}| \sim 0.5~\mu_{\rm B}$, induced by the SO coupling and further enhanced by on-site electron-electron interactions~\cite{PRL1998,PRB2014}. The inversion symmetry breaking gives rise to an effective Dzyaloshinskii-Moriya interaction, which is responsible for small canting of magnetic moments between different sublattices (\Fref{fig.H}). Contrary to regular Dzyaloshinskii-Moriya interactions~\cite{Dzyaloshinskii_weakF,Moriya_weakF}, caused by off-centrosymmetric displacements of intermediate ligand sites~\cite{Keffer}, these interactions are induced by antiferro orbital ordering. 

\par Results of numerical simulations in the magnetic field are summarized in~\Fref{fig.H}. The electric polarization was evaluated using Berry-phase theory, which was applied to the model Hamiltonian in the HF approximation~\cite{PRB2012}. Since threefold rotational symmetry is broken, $\vec{P}$ is allowed to have al three components. Here, we focus on the behavior of $P^{z}$, which dominates over $P^{x}$ and $P^{y}$~\cite{PRB2024}. The magnetic field dependence of spin and orbital magnetic moments is characterized by the hysteresis loop, where the field applied in the positive direction of $z$ leads to the saturation of magnetization, while in the opposite direction it leads to the reorientation of magnetization at around $H \sim -7~{\rm T}$. Corresponding $P^{z}$ has a characteristic butterflylike shape and undergoes the jump $\Delta P^{z} \sim 2.4~\mu{\rm C}/{\rm cm}^{2}$ at the point of reorientation of magnetization. This polarization jump is comparable with the one observed in CaBaCo$_4$O$_7$ ($\Delta P \sim 1.7~\mu{\rm C}/{\rm cm}^{2}$), which is believed to be the largest polarization change induced by magnetic field~\cite{CaBaCo4O7}. 

\section{\label{sec:summary} Summary and outlook}
\par First, let us summarize main principles for realizing orbitally induced FM-FE state, which we propose.

\par (i) The magnetic atoms should not be in the inversion centers. The atoms belonging to different sublattices can be connected by $\mathcal{I}$, and this $\mathcal{I}$ can be broken by antiferro orbital order. One promising example is the honeycomb lattice, considered in the persent work. Of course, there are other possibilities including corundum, some hexagonal and monoclinic structures. For instance, such a situation is realized in the $P6_{3}/mmc$ structure of BaCrO$_3$~\cite{BaCrO3} and $P2/c$ structure of CoWO$_4$~\cite{CoWO4}.

\par (ii) The $d^{2}$ configuration appears to be rather unique, at least for the octahedral environment. Two electrons are required to activate Hund's second rule and enforce the orbital degeneracy, so that the occupied orbitals could freely adjust their shape and minimizing the energy of exchange interactions via the orbital ordering. From this point of view, other configurations are less promising. The $d^{1}$ configuration can be excluded because single electron is not subjected to the Hund's rule physics. The level splitting in this case is controlled solely by the JT distortion, which lifts the orbital degeneracy, preventing the system from formation of antiferro orbital ordering. A similar situation occurs for the $d^{4}$ configuration in the high spin-state, where there is only one hole with the majority spin and no competition between orbitals states, which can be resolved by Hund's second rule. Furthermore, the $d^{4}$ configuration is subjected to strong JT distortion~\cite{Kanamori1960}. The $d^{3}$ configuration is not orbitally active because of large $10Dq$ splitting. Thus, considering less than half filled shell in octahedral environment, the only promising configuration seems to be $d^{2}$, which can be realized in Ti$^{2+}$, V$^{3+}$, and Cr$^{4+}$. 

\par (iii) The main parameters controlling the ground state of $d^{2}$ ions are $\Delta_{\rm tr}$, $B$, and $10Dq$. The crystal-field splitting $\Delta_{\rm tr}$ is caused by the JT distortion, which tends to lift the orbital degeneracy. The Racah parameter $B$ is responsible for Hund's second rule, which acts in the opposite direction, yielding the ground state with maximal orbital degeneracy. The octahedral splitting $10Dq$ does not alter the order of $t_{2g}$ levels directly. However, since Hund's interactions mix $e'_{g}$ and $e_{g}$ states, which are separated by $10Dq$, the strength of the Hund's rule coupling effectively depends on $10Dq$. Hund's second rule does not apply to isolated $t_{2g}$ manifold in the limit of large $10Dq$: the situation is similar to $p$ electron system, where in the high-spin state there is always only one possible orbital state. Therefore, the possibility of mixing of $e'_{g}$ and $e_{g}$ states is crucial for realizing Hund's second rule. While $B$ is not very sensitive to crystal environment, $\Delta_{\rm tr}$ and $10Dq$ strongly depend on details of the crystal structure and type of the ligand atoms. For our purposes, one would like to have smaller $\Delta_{\rm tr}$ and $10Dq$. Since they depend on hybridization between the transition-metal $d$ and ligand $p$ states, the I$^{-}$ ions seem to be more promising than O$^{2-}$ ones: the I $5p$ states are more diffuse, the V-I distance is significantly large and, therefore, the hybridization between V $3d$ and I $5p$ states is substantially weaker.

\par (iv) Another interesting option is the $d^{7}$ configuration of Co$^{2+}$ ions in the honeycomb lattice. Such materials are well know and widely discussed as potential candidates for realizing the Kitaev quantum spin liquid state~\cite{LiuKhaliullin,SanoKatoMotome}. The canonical examples are BaCo$_2$(AsO$_4$)$_2$~\cite{BaCo2As2O8} and Na$_2$Co$_2$TeO$_6$~\cite{Na2Co2TeO6}. The corresponding Kitaev model is formulated for the  Co$^{2+}$ ions having pseudospin-1/2 doublet in the ground state, which can be realized in the large $10Dq$ limit. However, in order to realize antiferro orbital order on the honeycomb lattice by activating Hund's second rule effects, we need the opposite situation, where $10Dq$ is relatively small. This will mix the pseudospin-1/2 doublet with other states. Thus, materials suitable for the FM-FE state are generally unsuitable for Kitaev physics.

\par To date, VI$_3$ remains the only potential candidate for realizing the orbitally induced FM-FE state, according to the theoretical proposal in~\cite{PRB2024}. The experimental situation remains rather controversial. VI$_3$ is indeed a ferromagnet with relatively high Curie temperature, $T_{\rm C} \approx 50$ K~\cite{VI3Kong}. Besides the FM transition, VI$_3$ exhibits at least two structural phase transitions, at around $78$ and $32$ K~\cite{VI3Dolezal,VI3Marchandier}. However, details of these transitions is a matter of dispute: there exist different proposals regarding crystal structure changes, the direction of these changes, as well as their interconnection with the magnetic properties of VI$_3$. For instance, according to several reports, the vanadium atoms across inversion centers in honeycomb layers become inequivalent, either structurally~\cite{VI3Son} or magnetically~\cite{VI3Gati}, meaning that these inversion centers are broken. All these data suggest that the crystal structure of VI$_3$ is rather fragile, which is consistent with our main idea that the orbital degrees of freedom in VI$_3$ remain active and the decision which orbitals become occupied and which remain empty may depend on tiny balance between several factors, including the experimental conditions. A more detailed discussion can be found in~\cite{PRB2024}.

\par It remains an open question whether the above principles (i)-(iv) are sufficient for realizing the FM-FE ground state in VI$_3$ and other materials. From the theoretical point of view, the orbital degeneracy is the very challenging problem, where various scenarios are possible~\cite{Khaliullin}. Nevertheless, we believe that the ferromagnetic ferroelectricity is one such plausible scenario, that should be considered alongside others.

\par Besides ferroelectric ferroelectricity, there are other fundamental issues related to the Hund's rule physics. The importance of Hund's first rule in the physics of strongly correlated materials is well recognized today~\cite{Georges}. The proper interaction describing these effects is $J_{\rm H}$. However, the effects related to Hund's second rule, driven by the Racah parameter $B$, remain largely unexplored. Due to the hierarchy of atomic Hund's rules, reflected in the condition $B \sim 0.1 J_{\rm H}$, exploring the effects of Hund's second rule effects requires working with a much smaller energy scale, which posses a challenge for numerical simulations. Nevertheless, implications of these effects to the physics of strongly correlated materials are rather interesting, as they provide a possibility for investigating new type of phenomena, such as ferroelectric ferroelectricity. They also provide a new look on other canonical problems. 

\par Particularly, orbital fluctuations were, and continue to be, one of the hot topic in the physics of $t_{2g}$ perovskite oxides, like LaVO$_3$ and YVO$_3$~\cite{Khaliullin}. One of the key questions is how well these fluctuations are quenched by the lattice distortions, which are typically present in these systems. The analysis of the crystal-field splitting, derived from electronic structure calculations, suggests that it can be quite strong~\cite{review2008}. However, this analysis was typically based on the model constructed separately for the $t_{2g}$ bands, which exclude any effects related to Hund's second rule. On the other hand, Hund's second rule is expected to play an important role in these two-electron systems and compete with the crystal-field splitting. Thus, the problem of orbital fluctuations in LaVO$_3$ and YVO$_3$ can be reconsidered on a new level, taking into account the effects of Hund's second rule in a more general model constructed simultaneously for the $t_{2g}$ and $e_{g}$ states. One of the key parameters of this model controlling the strength of effective interactions responsible for Hund's second rule will be $10Dq$. 

\par Another seminal problem is how to improve the exchange-correlation functional in electronic structure calculations in order to deal with the orbital magnetism. The density functional theory (DFT), providing general foundation for modern electronic structure calculations, is formally exact. However, to be practically applicable, it is supplemented with additional approximations, such as local spin density approximation (LSDA) or generalized gradient approximation (GGA), which largely rely on the model of electron gas and fail to account properly for the physics of intraatomic interactions, including those responsible for Hund's second rule. Very often, the orbital magnetization calculated within LSDA or GGA is severely underestimated. In order to correct the orbital magnetization, one strategy was to mimic the effects of Hund's second rule in electronic structure calculations, by adding to the total energy a phenomenological term, which would (i) be proportional to corresponding Racah parameter $B$ and (ii) tend to enhance $\boldsymbol{M}_{\rm L}$~\cite{Brooks1989,Norman1990,Norman1991}. The typical choice is $-B\boldsymbol{M}_{\rm L}^2$~\cite{Brooks1989,Norman1990}, though there are also other suggestions~\cite{Norman1991}. Another contribution to the orbital magnetization is driven by the on-site Coulomb repulsion $U$~\cite{PRL1998,Terakura1984,PRB2014}. It does not contribute to the energy of isolated atoms with the given integer number of electrons~\cite{PRB94}, but emerges in solids, where the atom behaves as an open electron system and can freely exchange electrons with other atoms. The key question in this context is how to properly evaluate the screening of $U$, especially in metallic systems~\cite{PRL2005}. 

\par By treating $U$ as an adjustable parameter, as is frequently done in electronic structure calculations, one can easily reproduce the experimental magnetization. However, it does not bring us closer to microscopic understanding of this problem because there can be other interactions controlling the value of $\boldsymbol{M}_{\rm L}$. Particularly, what is the relative importance of on-site Coulomb repulsion $U$ and interactions responsible for Hund's second rule? In this respect, one should clearly understand that the proposed correction $-B\boldsymbol{M}_{\rm L}^2$ is totally empirical and does not reflect the true physics of atomic Hund's rules. For instance, it will enhance $\boldsymbol{M}_{\rm L}$ even when there is only one electron, i.e. when Hund's rules are not operative. However, it does not mean that Hund's rule interactions cannot contribute to $\boldsymbol{M}_{\rm L}$ in principle. A better correction can be obtained considering intraatomic exchange interaction energy in the HF approximation~\cite{PRL1998,PRB2014}. The applications can be also rather interesting. For instance, one can expect that the orbital magnetization in $d^{1}$ and $d^{2}$ materials should behave differently: in the former case, it is controlled solely by Coulomb $U$, while in the latter case there will be an additional contribution associated with Hund's second rule.

\ack I am indebted to Ryota Ono and Sergey Nikolaev for collaboration on earlier stages of this project~\cite{PRB2024}. MANA is supported by World Premier International Research Center Initiative (WPI), MEXT, Japan.

\end{document}